\documentclass{article}

    \PassOptionsToPackage{numbers, compress}{natbib}

\usepackage[final]{neurips_2024}
\usepackage{graphicx}    
\usepackage{amsmath}     
\usepackage{algorithm}   
\usepackage{algpseudocode}
\usepackage{subfigure}   
\usepackage{booktabs}    
\usepackage{xcolor}      
\usepackage{amsfonts}    
\usepackage{nicefrac}    
\usepackage{hyperref}    
\usepackage{url}         
\usepackage{wrapfig}     
\usepackage{multirow}    
\usepackage{newunicodechar} 
\usepackage[capitalize,noabbrev]{cleveref} 
\usepackage{enumitem}
\usepackage{multicol}
\usepackage{adjustbox}
\usepackage{float}

\usepackage[utf8]{inputenc} 
\usepackage[T1]{fontenc}    
\usepackage{hyperref}       
\usepackage{url}            
\usepackage{booktabs}       
\usepackage{amsfonts}       
\usepackage{nicefrac}       
\usepackage{microtype}      
\usepackage{xcolor}         
\newcommand{\header}[1]{\textbf{#1}}

\title{TTT-UNet: Enhancing U-Net with Test-Time Training Layers for Biomedical Image Segmentation}

\author{
  $\text{\textbf{Rong Zhou}}^{\textbf{1}\thanks{\raggedright Equal contribution. Contact emails: \texttt{roz322@lehigh.edu, zyuan2@nd.edu, zhy423@lehigh.edu, kennysweson@gmail.com.}}}$~~~~~~~~
  $\text{\textbf{Zhengqing Yuan}}^{\textbf{3}*}$~~~~~~~~~~
  $\text{\textbf{Zhiling Yan}}^{\textbf{1}*}$~~~~~~~~~~\\
  $\text{\textbf{Weixiang Sun}}^{\textbf{2}*}$~~~~~~~~~~
  $\text{\textbf{Kai Zhang}}^{\textbf{1}}$~~~~~~~~~~
  $\text{\textbf{Yiwei Li}}^{\textbf{4}}$~~~~~~~~~~\\
  $\text{\textbf{Yanfang Ye}}^{\textbf{3}}$~~~~~~~~~~
  $\text{\textbf{Xiang Li}}^{\textbf{5}}$~~~~~~~~~~
  $\text{\textbf{Lifang He}}^{\textbf{1}}$~~~~~~~~~~
  $\text{\textbf{Lichao Sun}}^{\textbf{1}}$\\
  ${}^{\textbf{1}}$Lehigh University, 
  ${}^{\textbf{2}}$Northeastern University, 
  ${}^{\textbf{3}}$University of Notre Dame, \\
  ${}^{\textbf{4}}$University of Georgia, 
  ${}^{\textbf{5}}$Massachusetts General Hospital and Harvard Medical School\\
}
\begin{document}

\maketitle

\begin{abstract}
  Biomedical image segmentation is crucial for accurately diagnosing and analyzing various diseases. However, Convolutional Neural Networks (CNNs) and Transformers, the most commonly used architectures for this task, struggle to effectively capture long-range dependencies due to the inherent locality of CNNs and the computational complexity of Transformers. To address this limitation, we introduce TTT-UNet, a novel framework that integrates Test-Time Training (TTT) layers into the traditional U-Net architecture for biomedical image segmentation. TTT-UNet dynamically adjusts model parameters during the testing time, enhancing the model's ability to capture both local and long-range features. We evaluate TTT-UNet on multiple medical imaging datasets, including 3D abdominal organ segmentation in CT and MR images, instrument segmentation in endoscopy images, and cell segmentation in microscopy images. The results demonstrate that TTT-UNet consistently outperforms state-of-the-art CNN-based and Transformer-based segmentation models across all tasks. The code is available at \url{https://github.com/rongzhou7/TTT-UNet}.
\end{abstract}

\section{Introduction}
Image segmentation plays a crucial role in medical imaging, as it empowers medical professionals to identify biological structures and measure their morphology, aiding in the analysis and diagnosis of various diseases~\cite{qureshi2023medical, cao2023comprehensive}. In recent years, convolutional neural networks (CNNs)~\cite{lecun1995convolutional} have emerged as a promising approach in the field of biomedical image segmentation. Among various CNN-based techniques, U-Net~\cite{ronneberger2015u} stands out for its straightforward structure and significant adaptability. Many enhancements and iterations~\cite{huang2020unet, cao2022swin, zhou2018unet++, zhou2019unet++, hatamizadeh2022unetr, hatamizadeh2021swin} have been developed based on this U-shaped architecture, typically featuring a symmetric encoder-decoder design to capture multi-scale image features through convolutional operations. Leveraging this foundation, significant advancements have been achieved across a wide range of medical imaging applications~\cite{yan2024biomedical, zhang2024generalist, chen2024ma, sun2024bora, liu2024sora, zheng2024llm4brain, zhou2023attentive, zhou2023integrating}. These include cardiac segmentation in magnetic resonance (MR) imaging~\cite{wang2021sk}, multi-organ delineation in computed tomography (CT) scans~\cite{li2018h}, and others~\cite{hasan2019u, safarov2021denseunet}.

Despite the remarkable representational capabilities of CNN-based models, their architectural design exhibits an inherent limitation in modeling long-range dependencies within images, because convolutional kernels are inherently local~\cite{chen20233d}. While skip connections in the U-Net architecture facilitate the merging of low-level details with high-level features, they mainly serve to directly merge local features, which does not substantially boost the network's ability to model long-range dependencies. This limitation becomes especially evident in scenarios with large inter-patient variations in shape, size, etc~\cite{chen20233d}. Such variability poses challenges to the CNN framework's ability to consistently and accurately capture information across extended spatial contexts, highlighting the need for innovative approaches to address this fundamental constraint.

Recognizing the limitations of CNNs in capturing long-range dependencies, the research community has shifted interest towards Transformer models for their ability to naturally understand global contexts~\cite{ji2021improving}. 
This transition is evidenced in biomedical image segmentation, where approaches like TransUNet~\cite{chen2021transunet}, UNETR~\cite{hatamizadeh2022unetr},  SwinUNETR~\cite{hatamizadeh2021swin} demonstrate the potential of integrating Transformers. 
These hybrid models that blend CNNs for high-resolution spatial detail and Transformers for global context emerge as a more effective strategy. 

Despite their ability to capture global dependencies, Transformers are computationally intensive~\cite{gu2023mamba}, especially in dense biomedical image segmentation tasks. 
Mamba~\cite{gu2023mamba}, a state-space model designed for efficient sequence modeling, offers a more computationally efficient approach to long-range dependency modeling. Building on this, U-Mamba~\cite{ma2024u} integrates Mamba within U-Net, effectively combining high-resolution spatial detail with long-range dependency modeling to enhance biomedical image segmentation. Despite these advancements, U-Mamba and similar models, still face challenges in expressiveness, particularly over extended contexts, where their fixed-size hidden states limit their ability to capture complex and nuanced dependencies. 

Recently, TTT (Test-Time Training)~\cite{sun2024learning} have emerged as a new class of sequence modeling layers with linear complexity and an expressive hidden state. TTT treats the traditional fixed hidden state as a machine learning model itself, which can be dynamically updated through self-supervised learning. This dynamic adjustment allows the model to refine its parameters based on test data, providing greater flexibility and expressiveness in capturing intricate long-range dependencies.
In comparison to Transformers and Mamba, TTT layers not only maintain efficiency but also offer superior performance in handling long-context sequences.

In this paper, we introduce TTT-UNet, a novel hybrid architecture that incorporates TTT layers within the traditional U-Net framework to address the inherent limitations in modeling long-range dependencies in biomedical image segmentation tasks. 
The TTT layers dynamically adapt its parameters during test time, allowing it to more effectively capture both localized details and long-range dependencies.
Our extensive experiments across various medical imaging datasets demonstrate that TTT-UNet consistently outperforms existing state-of-the-art models. The results highlight the model's effectiveness in handling complex anatomical structures and its robustness in diverse clinical scenarios. Particularly, TTT-UNet has shown significant improvements in biomedical image segmentation tasks, making it a versatile solution for medical image analysis.
Our contributions are summarized as follows:
\begin{itemize}
\item 

We introduce TTT-UNet, an enhanced U-Net architecture integrated with TTT layers, which allows the model to perform self-supervised adaptation during test time. This hybrid design effectively tackles the challenge of modeling long-range dependencies and improves the model's generalization capability across diverse data distributions.

\item 
TTT-UNet has been rigorously evaluated on a diverse set of medical imaging datasets, including 3D abdominal organ segmentation in CT and MRI scans, instrument segmentation in endoscopy images, and cell segmentation in microscopy images. The results demonstrate consistent improvements over state-of-the-art models in both 3D and 2D segmentation. 

\end{itemize}
In summary, TTT-UNet represents a significant advancement in biomedical image segmentation, offering a robust and adaptable approach that leverages the strengths of CNNs and TTT layers. This work lays the foundation for future developments in adaptive and context-aware medical image analysis technologies.

\section{Related work}
\subsection{U-Net and variants}
CNN-based and Transformer-based models have significantly advanced the field of biomedical image segmentation. U-Net~\cite{ronneberger2015u}, a representative among CNN-based approaches, features a symmetrical encoder-decoder architecture enhanced with skip connections to better preserve details. Various enhancements~\cite{myronenko20193d}, such as the self-configuring nnU-Net~\cite{isensee2021nnu} framework, have been built on this U-shaped design, demonstrating robust performance across a variety of biomedical image segmentation challenges. For Transformer, TransUnet~\cite{chen2021transunet} stands out by integrating the Vision Transformer (ViT)~\cite{dosovitskiy2020image} for feature extraction in the encoding phase and coupling it with CNN for decoding, demonstrating its capability for processing global information. Swin-UNETR~\cite{hatamizadeh2021swin} and UNETR~\cite{hatamizadeh2022unetr} blend Transformer architectures with traditional U-Net to enhance 3D imaging analysis. Additionally, Swin-UNet~\cite{cao2022swin} delves into the use of Swin Vision Transformer blocks~\cite{liu2021swin} within a U-Net framework, further expanding the exploration of Transformer technology in medical imaging.

\subsection{Hybrid models}

SSMs, such as Mamba, have recently gained prominence as a powerful component for developing deep networks, achieving cutting-edge performance in analyzing long-sequence data~\cite{goel2022s, fu2022hungry}. In the realm of biomedical image segmentation, U-Mamba~\cite{ma2024u} presents a novel SSM-CNN hybrid approach, signifying the first application of SSMs in the medical image domain. Further developments include SegMamba~\cite{xing2024segmamba} and nnMamba~\cite{gong2024nnmamba}, which combine SSMs in the encoder with CNNs in the decoder, illustrating the versatility and effectiveness of SSMs in enhancing medical imaging analysis.

\section{Method}
\begin{figure}[ht]
    \centering
\includegraphics[width=\linewidth]{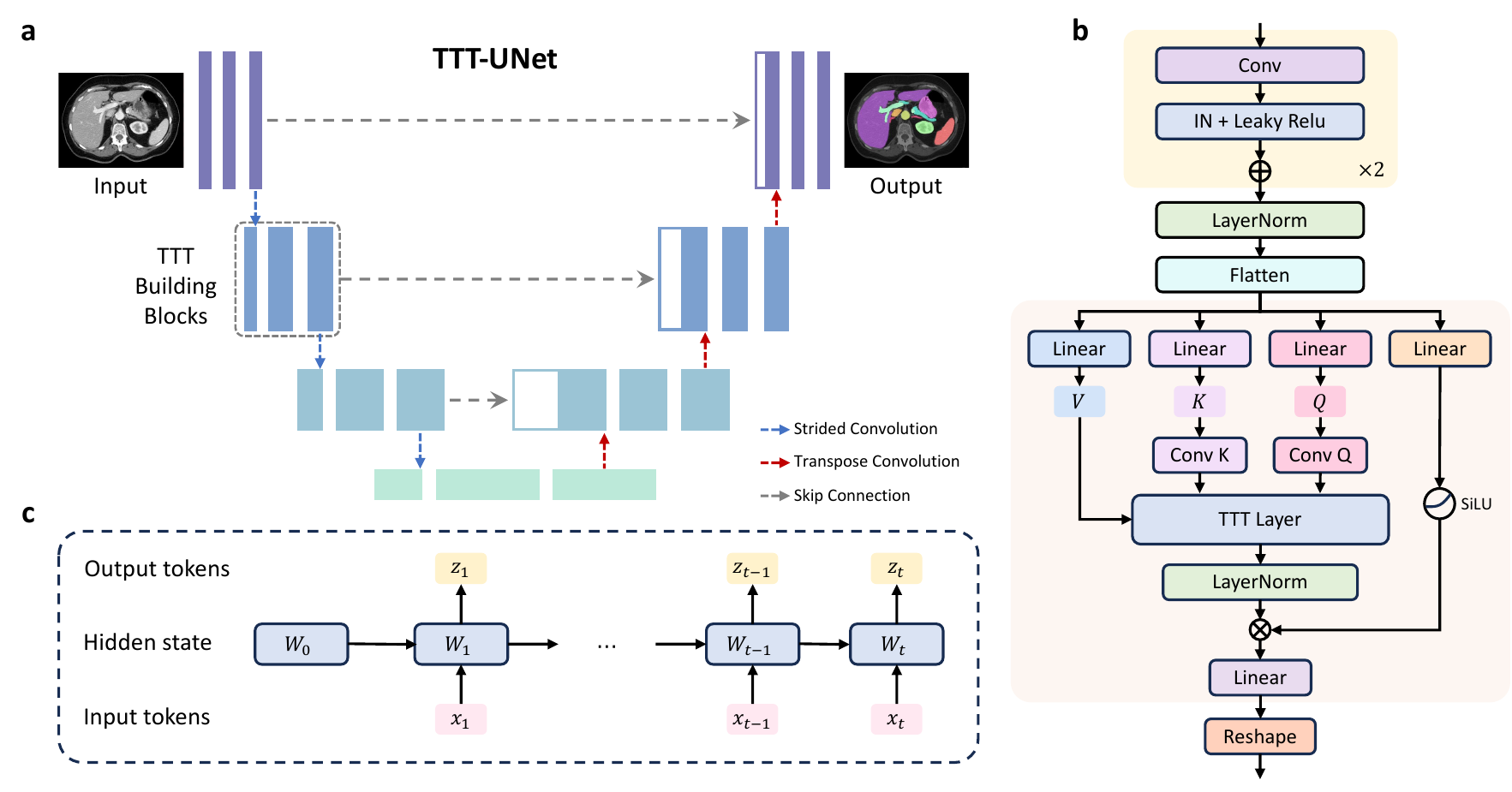}
    \caption{ \textbf{(a)} The overall framework of TTT-UNet. \textbf{(b)} TTT Building Block. \textbf{(c)} TTT Layer. }
    \label{fig:frame}
\end{figure}

TTT-UNet follows the conventional U-Net structure, designed to effectively capture both local features and long-range dependencies. 
As shown in Figure~\ref{fig:frame}, TTT-UNet integrates Test-Time Training (TTT) layers into the Mamba blocks within the U-Net network. 
This integration enables the model to continuously update its parameters based on test data, enhancing its feature extraction capabilities in the encoder and allowing it to adaptively learn long-range dependencies. 
Subsequently, we introduce the TTT layer and then describe how it is integrated into the Mamba blocks within the U-Net architecture.

\subsection{TTT layers} \label{TTTLayer}

Traditional sequence modeling layers, such as RNNs, compress the context of a sequence $x_1, \dots, x_t$ into a fixed-size hidden state \(h_t\). For RNNs, the hidden state \(h_t\) at time step \(t\) is updated based on the current input \(x_t\) and the previous hidden state \(h_{t-1}\) through linear transformation matrices $\theta_h$ and $\theta_x$ and a non-linear activation function \(\sigma\):
\[
h_t = \sigma(\theta_h h_{t-1} + \theta_x x_t),
\]
where \(\theta_h\) and \(\theta_x\) are learned parameters. The output \(z_t\) is then generated from the hidden state:
\[
z_t = \phi(h_t),
\]
where \(\phi\) represents a linear or non-linear transformation.

However, the fixed size of the hidden state limits performance when dealing with long contexts due to its finite capacity to represent contextual information.
To address this limitation, a new class of sequence modeling layers, referred to as TTT layers \cite{sun2024learning} is introduced, where the hidden state is treated as a trainable model and is updated through self-supervised learning.




Specifically, in a TTT layer (\textbf{Fig. \mbox{\ref{fig:frame}c}})
, the hidden state \( h_t \) at time step \( t \) is treated as a trainable model \( f \) with weights \( W_t \), which is updated based on the current input \( x_t \):
\[
W_t = W_{t-1} - \eta \nabla \ell(W_{t-1}; x_t)
\]
The output token \(z_t\) is then generated using trainable model \( f \) with weights \( W_t \):
\[
z_t = f(x_t; W_t)
\]
In the basic naive version, the self-supervised loss $\ell$ aims to reconstruct the corrupted input \( \tilde{x}_t \). This approach is straightforward and focuses on learning to recover the original input from its corrupted version:
\[
\ell(W; x_t) = \|f(\tilde{x}_t; W) - x_t\|^2
\]
While this naive reconstruction method is effective in certain scenarios, it has inherent limitations in capturing the complex dependencies within the input data, especially in tasks requiring a more nuanced understanding of the input context.

To address these limitations, we follow a more sophisticated self-supervised task that leverages multiple views of the input data. Instead of directly reconstructing the corrupted input, we introduce learnable matrices \( \theta_K \) and \( \theta_V \) to project the input into different views. The \textit{training view} \( K=\theta_K x_t \) captures the essential information needed for learning, while the \textit{label view} \( V=\theta_V x_t \) provides a target for reconstruction:
\[
\ell(W; x_t) = \|f(\theta_K x_t; W) - \theta_V x_t\|^2
\]
This approach allows the model to selectively focus on the most relevant features of the input, improving its ability to capture long-range dependencies and subtle relationships within the data. 

The output token \(z_t\) is then generated as follow:
\[
z_t = f(\theta_Q x_t; W_t), 
\]
where \(f\) is a function parameterized by \(W_t\), which can be a linear model or a multi-layer perceptron (MLP).
Here, the projection \( \theta_Q \) is used to obtain the test view \(Q= \theta_Q x_t \), which introduces additional flexibility by allowing the model to emphasize different aspects of the input data during inference. This approach enables the model to focus on the most informative features in the context of the current test case, thereby enhancing its ability to adapt to new, unseen data.

\subsection{TTT-UNet architecture}

As shown in \textbf{Fig. \mbox{\ref{fig:frame}}}, the TTT-UNet architecture integrates the traditional U-Net structure with TTT layers, allowing the network to adapt during testing through self-supervised learning dynamically. 
The architecture is composed of an encoder-decoder structure, where the encoder is enhanced with TTT building blocks to improve adaptability, while the decoder follows the standard U-Net design focused on reconstructing the segmentation map.

\header{Encoder.} The encoder in TTT-UNet follows the traditional U-Net design, comprising multiple convolutional layers. These layers are interspersed with TTT building blocks, which are critical components that enable the model to adjust its parameters dynamically during test time. 
Each layer in the encoder progressively downscales the input image while capturing both local and long-range features essential for segmentation tasks. 
Including TTT building blocks within the encoder ensures the model can adapt to varying data distributions encountered during testing.

\header{TTT building blocks.} 
As illustrated in \textbf{Fig. \mbox{\ref{fig:frame}b}}, the TTT building blocks are the core components that allow for the test-time adaptability of the model.
Initially, the input features pass through two successive Residual blocks \cite{he2016deep}, each comprising a standard convolutional layer, followed by Instance Normalization (IN) \cite{ulyanov2016instance} and Leaky ReLU activation \cite{maas2013rectifier}. 
Subsequently, the features are normalized using Layer Normalization \cite{ba2016layer}, and flattened, making them suitable for linear transformations. 
Then the flattened features undergo three separate linear transformation branches, obtaining the different features denoted as  \(V\), \(K\), and \(Q\) respectively. 
Additional convolutional operations (\texttt{Conv} \(K\) and \texttt{Conv} \(Q\)) are applied to the \(K\) and \(Q\) vectors, allowing the model to focus on specific aspects of the features during test-time training. 
Meanwhile, the fourth branch performs a linear transformation followed by a SiLU activation function \cite{hendrycks2016gaussian}, further enriching the feature representations.
Then the processed  \(V\), \(K\), and \(Q\) are fed into the TTT Layer, where self-supervised learning occurs. In this layer, the model dynamically updates its weights based on the self-supervised task applied to the processed \(V\), \(K\), and \(Q\) vectors, as detailed in \ref{TTTLayer}. The output from the TTT Layer is further normalized using Layer Normalization \cite{ba2016layer} before being passed on.
Finally, this output and the fourth branch output mentioned before are combined via the Hadamard product, followed by a linear transformation and reshaping to fit the required dimensions for subsequent layers in Decoder.


\header{Decoder} The decoder in our model maintains the classic U-Net structure, integrating Residual blocks and transposed convolutions to enhance the capture of detailed local features and support resolution recovery. We also incorporate the skip connections in U-Net, ensuring the effective transfer of hierarchical features from the encoder to the decoder. The final output of the decoder is refined through a 1 × 1 × 1 convolutional layer and a Softmax activation, which generates the final segmentation probability map.

Furthermore, we implement two variants of TTT-UNet: TTT-UNet\_Bot and TTT-UNet\_Enc. In TTT-UNet\_Bot, TTT layers are applied solely in the bottleneck, while the rest of the architecture consists of standard Residual blocks. 
In TTT-UNet\_Enc, TTT layers are incorporated throughout the encoder, allowing for a broader integration of self-supervised adaptation.



\section{Experiments}
\subsection{Datasets}
To evaluate the performance and scalability of TTT-UNet, we utilize four biomedical image datasets across a variety of segmentation tasks and imaging modalities, including Abdomen CT dataset~\cite{ma2023unleashing}, Abdomen MRI dataset~\cite{ji2022amos}, Endoscopy dataset~\cite{allan20192017} and Microscopy dataset~\cite{ma2023multi}. The basic information of these datasets is summarized in Table~\ref{tab:datasets}.

\begin{table}[ht]
\centering
\caption{Dataset information.}
\begin{tabular}{lcccc}
\toprule
\textbf{Dataset}    &     \textbf{Dimension}  & \textbf{\#Training Image}  & \textbf{\#Testing Image}  & \textbf{\#Targets}  \\ 
\midrule
Abdomen CT               & 3D                 & 50 (4794 slices)           & 50 (10894 slices)        & 13                 \\ 
Abdomen MRI              & 3D                 & 60 (5615 slices)           & 50 (3357 slices)         & 13                 \\ 
Endoscopy images         & 2D                 & 1800                       & 1200                     & 7                  \\ 
Microscopy images        & 2D                 & 1000                       & 101                      & 2                  \\ 
\bottomrule
\end{tabular}
\label{tab:datasets}
\end{table}


\textbf{Abdomen CT.}
The Abdomen CT~\cite{ma2023unleashing} dataset, from the MICCAI 2022 FLARE challenge, includes the segmentation of 13 abdominal organs from 50 CT scans in both the training and testing sets. The organs segmented include the liver, spleen, pancreas, kidneys, stomach, gallbladder, esophagus, aorta, inferior vena cava, adrenal glands, and duodenum.

\textbf{Abdomen MRI.}
The Abdomen MR~\cite{ji2022amos} dataset, from the MICCAI 2022 AMOS Challenge, focuses on the segmentation of the same 13 abdominal organs, using MRI scans. It consists of 60 MRI scans for training and 50 for testing. Additionally, we generate a 2D version of this dataset by converting the 3D abdominal MRI scans into 2D slices. This conversion enables us to evaluate TTT-UNet under the common 2D segmentation setting, which is widely used in practice due to its lower computational requirements. The conversion retains the same 13 organs, ensuring consistent evaluation across both 2D and 3D modalities.

\textbf{Endoscopy images.} 
 From the MICCAI 2017 EndoVis Challenge~\cite{allan20192017}, this dataset focuses on instrument segmentation within endoscopy images, featuring seven distinct instruments, including the large needle driver, prograsp forceps, monopolar curved scissors, cadiere forceps, bipolar forceps, vessel sealer, and a drop-in ultrasound probe. The dataset is split into 1800 training frames and 1200 testing frames.

\textbf{Microscopy images.} 
This dataset, from the NeurIPS 2022 Cell Segmentation Challenge~\cite{ma2023multi}, is used for cell segmentation in microscopy images, consisting of 1000 training images and 101 testing images. Following U-Mamba~\cite{ma2024u}, we address this as a semantic segmentation task, focusing on cell boundaries and interiors rather than instance segmentation.

\subsection{Experimental setup}

\begin{table}[ht]
\centering
\caption{TTT-UNet configurations for each dataset.}
\begin{tabular}{lcccc}
\toprule
\textbf{Configurations}  & \textbf{Patch size}   & \textbf{Batch size} & \textbf{\# Stages} & \textbf{\# Pooling per axis} \\ 
\midrule
Abdomen CT               & (40, 224, 192)        & 2                   & 6                 & (3, 3, 5)                    \\ 
3D Abdomen MR            & (48, 160, 224)        & 2                   & 6                 & (3, 5, 5)                    \\ 
2D Abdomen MR            & (320, 320)            & 30                  & 7                 & (6, 6)                       \\ 
Endoscopy                & (384, 640)            & 13                  & 7                 & (6, 6)                       \\ 
Microscopy               & (512, 512)            & 12                  & 8                 & (7, 7)                       \\ 
\bottomrule
\end{tabular}
\label{tab:configs}
\end{table}

The setting of our experiments is the same as that in U-Mamba~\cite{ma2024u} and nnU-Net~\cite{isensee2021nnu} to ensure a fair comparison, as shown in Table~\ref{tab:configs}
We adopt an unweighted combination of Dice loss and cross-entropy loss for all datasets and utilize the SGD optimizer with an initial learning rate of 1e-2. The training duration for each dataset is set to 1000 epochs, conducted on a single NVIDIA A100 GPU. Leveraging the self-configuring capabilities from nnU-Net, the number of network blocks adjusts automatically according to the dataset.
For evaluation metrics, we employ the Dice Similarity Coefficient (DSC) and Normalized Surface Distance (NSD) to assess performance in 
abdominal multi-organ segmentation for MR scans, as well as 
instrument segmentation in Endoscopy images. For the cell segmentation task, we utilize the F1 score to evaluate method performance. 

\subsection{Baselines and metrics}
In our evaluation of TTT-UNet, we compare against two prominent CNN-based segmentation methods: nnU-NET~\cite{isensee2021nnu} and SegResNet~\cite{myronenko20193d}. Additionally, we include a comparison with UNETR~\cite{hatamizadeh2022unetr} and SwinUNETR~\cite{hatamizadeh2021swin}, a Transformer-based network that has gained popularity in biomedical image segmentation tasks. U-Mamba~\cite{ma2024u}, a recent method based on the Mamba model, is also included in our comparison to provide a comprehensive overview of its performance. For each model, we implement their recommended optimizers to ensure consistency in training conditions. To maintain fairness across all comparisons, we apply the default image preprocessing in nnU-NET~\cite{isensee2021nnu}.

\subsection{Quantitative segmentation results}

\begin{table}[htb]
\caption{Results summary of 2D segmentation tasks: organ segmentation in abdomen MRI scans, instruments segmentation in endoscopy images, and cell segmentation in microscopy images.}\label{tab:results-2d}
\centering
\begin{adjustbox}{width=0.99\textwidth}
\begin{tabular}{lcc|ll|c}
\toprule
\multirow{2}{*}{Methods} & \multicolumn{2}{c|}{Organs in Abdomem MRI}              & \multicolumn{2}{c|}{Instruments in Endoscopy}           & Cells in Microscopy        \\ \cline{2-6} 
                         & DSC                        & NSD                        & \multicolumn{1}{c}{DSC}    & \multicolumn{1}{c|}{NSD}   & F1                         \\ \hline
nnU-Net                  & 0.7450$\pm$0.1117          & 0.8153$\pm$0.1145          & 0.6264$\pm$0.3024          & 0.6412$\pm$0.3074          & 0.5383$\pm$0.2657          \\
SegResNet                & 0.7317$\pm$0.1379          & 0.8034$\pm$0.1386          & 0.5820$\pm$0.3268          & 0.5968$\pm$0.3303          & 0.5411$\pm$0.2633          \\
UNETR                    & 0.5747$\pm$0.1672          & 0.6309$\pm$0.1858          & 0.5017$\pm$0.3201          & 0.5168$\pm$0.3235          & 0.4357$\pm$0.2572          \\
SwinUNETR                & 0.7028$\pm$0.1348          & 0.7669$\pm$0.1442          & 0.5528$\pm$0.3089          & 0.5683$\pm$0.3123          & 0.3967$\pm$0.2621          \\ 
U-Mamba\_Bot             & 0.7588$\pm$0.1051 & 0.8285$\pm$0.1074 & 0.6540$\pm$0.3008 & 0.6692$\pm$0.3050 & 0.5389$\pm$0.2817 \\
U-Mamba\_Enc             & 0.7625$\pm$0.1082 & 0.8327$\pm$0.1087 & 0.6303$\pm$0.3067 & 0.6451$\pm$0.3104 & 0.5607$\pm$0.2784 \\ \hline
TTT-UNet\_Bot &\textbf{0.7750$\pm$0.1022} & \textbf{0.8452$\pm$0.1080}&\textbf{0.6643$\pm$0.3018} & \textbf{0.6799$\pm$0.3056} & \textbf{0.5818$\pm$0.2410} \\
TTT-UNet\_Enc & \textbf{0.7725$\pm$0.1044} & \textbf{0.8540$\pm$0.1032} &\textbf{0.6696$\pm$0.3018} & \textbf{0.6820$\pm$0.3080} & \textbf{0.5773$\pm$0.2435} \\
\hline
\end{tabular}
\end{adjustbox}
\end{table}

Table~\ref{tab:results-2d} presents the results of 2D segmentation tasks, comparing the performance of various models across three datasets: organ segmentation in Abdomen MRI, instrument segmentation in endoscopy images, and cell segmentation in microscopy images. 
For the organ segmentation task in Abdomen MRI, TTT-UNet models significantly outperformed other methods. The TTT-UNet\_Bot variant achieved the highest DSC of 0.7750±0.1022 and NSD of 0.8452±0.1080, while TTT-UNet\_Enc closely followed with a DSC of 0.7725±0.1044 and an NSD of 0.8540±0.1032. These results suggest that the TTT-UNet's ability to adapt its parameters at test time provides a considerable advantage in accurately segmenting organs, where anatomical variability is common. This adaptability likely contributes to the model's superior performance, enabling it to better generalize across different patient scans and handle the complexities of MRI data.

In the instrument segmentation task for endoscopy images, TTT-UNet variants again demonstrated superior performance, with TTT-UNet\_Bot achieving a DSC of 0.6643±0.3018 and NSD of 0.6799±0.3056. The TTT-UNet\_Enc model further improved these metrics, with a DSC of 0.6696±0.3018 and NSD of 0.6820±0.3080. These results indicate that the TTT layers within the model effectively capture the fine details of surgical instruments, which are often challenging to segment due to their small size and variability in appearance. The test-time adaptation provided by TTT layers helps in refining the segmentation boundaries, making the model more precise in instrument delineation.

For the cell segmentation task in microscopy images, the TTT-UNet models once again outperformed their counterparts. The TTT-UNet\_Bot achieved the highest F1 score of 0.5818±0.2410, followed by TTT-UNet\_Enc with an F1 score of 0.5773±0.2435. The superior performance in this task highlights the robustness of TTT-UNet in handling high variability and noise in microscopy data. The ability to dynamically adjust to new test samples allows the model to focus on relevant features, thereby improving segmentation accuracy even in challenging scenarios like cell segmentation.

\begin{table}[htb]
\caption{Results summary of 3D organ segmentation on abdomen CT and MRI datasets.}\label{tab:results-3d}
\centering
\begin{adjustbox}{width=0.99\textwidth}
\begin{tabular}{lcc|cc}
\toprule
\multirow{2}{*}{Methods} & \multicolumn{2}{c|}{Organs in Abdomen CT}                 & \multicolumn{2}{c}{Organs in Abdomen MRI}                 \\ \cline{2-5} 
                         & DSC                    & NSD                    & DSC                    & NSD                    \\ \hline
nnU-Net                  & 0.8615$\pm$0.0790          & 0.8972$\pm$0.0824          & 0.8309$\pm$0.0769          & 0.8996$\pm$0.0729          \\
SegResNet                & 0.7927$\pm$0.1162          & 0.8257$\pm$0.1194          & 0.8146$\pm$0.0959          & 0.8841$\pm$0.0917          \\
UNETR                    & 0.6824$\pm$0.1506          & 0.7004$\pm$0.1577          & 0.6867$\pm$0.1488          & 0.7440$\pm$0.1627          \\
SwinUNETR    & 0.7594$\pm$0.1095 & 0.7663$\pm$0.1190 & 0.7565$\pm$0.1394          & 0.8218$\pm$0.1409  \\

U-Mamba\_Bot             & 0.8683$\pm$0.0808 & \textbf{0.9049$\pm$0.0821} & 0.8453$\pm$0.0673 & 0.9121$\pm$0.0634 \\
U-Mamba\_Enc             & 0.8638$\pm$0.0908 & 0.8980$\pm$0.0921 & 0.8501$\pm$0.0732 & 0.9171$\pm$0.0689 \\
\hline
TTT-UNet\_Bot & \textbf{0.8709}$\pm$\textbf{0.1011} & 0.8995$\pm$0.0721  & \textbf{0.8677}$\pm$\textbf{0.0482}  & \textbf{0.9247}$\pm$\textbf{0.0631}\\  

\hline
\end{tabular}
\end{adjustbox}
\end{table}

Table~\ref{tab:results-3d} presents the results of 3D organ segmentation tasks on the Abdomen CT and Abdomen MRI datasets, comparing the performance of several state-of-the-art models. For the Abdomen CT dataset, TTT-UNet\_Bot achieves the highest DSC score of 0.8709±0.1011, slightly outperforming U-Mamba\_Bot (0.8683±0.0808) and U-Mamba\_Enc (0.8638±0.0908). The small variance in TTT-UNet\_Bot's performance suggests that the model is not only accurate but also consistent across different test samples. Additionally, the NSD score of 0.8995±0.0721 for TTT-UNet\_Bot further supports its ability to preserve organ shapes and boundaries effectively.

For the Abdomen MRI dataset, TTT-UNet\_Bot again demonstrates superior performance, achieving a DSC of 0.8677±0.0482 and an NSD of 0.9247±0.0631, outperforming all other models. The small variance in the DSC and NSD scores highlights the robustness of TTT-UNet\_Bot, suggesting that it generalizes well across different MRI samples, even in the presence of anatomical variability and challenging contrasts.

Overall, TTT-UNet\_Bot consistently outperforms other models in both CT and MRI segmentation tasks, not only achieving higher mean performance scores but also demonstrating lower variance, which indicates stable and reliable segmentation results. The test-time adaptation enabled by TTT layers plays a key role in this enhanced performance, allowing the model to effectively handle the complexities of 3D biomedical image segmentation.

\subsection{Qualitative segmentation results}

\begin{figure}[ht]
    \centering
    \includegraphics[width=\linewidth]{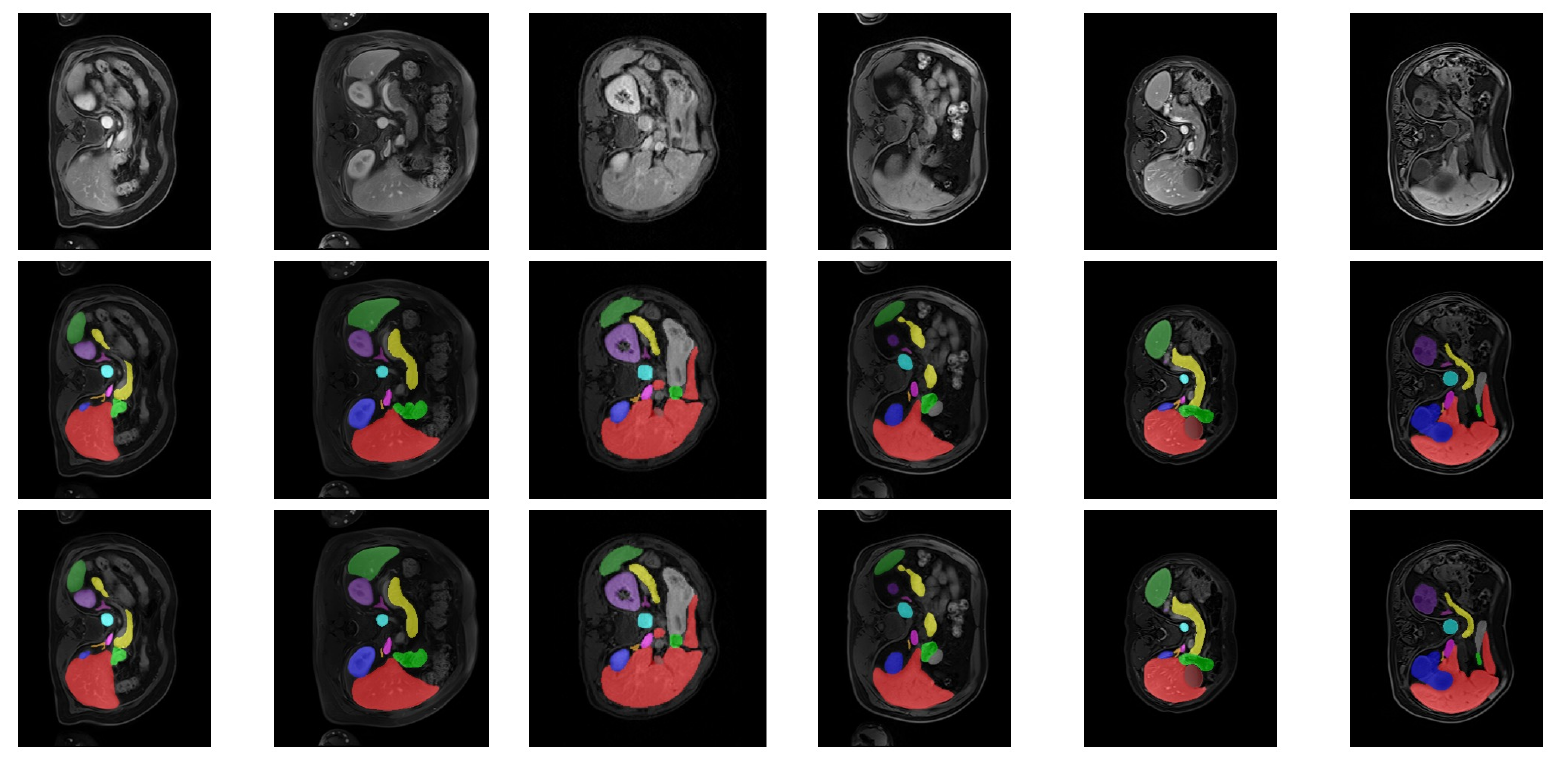}
    \vspace{-15pt}
    \caption{The visualization results of TTT-UNet on Abdomen MRI datasets. The first row shows the original images, the middle row shows the ground truth, and the bottom row shows the TTT-UNet predictions.}
    \label{fig: vismir}
    \vspace{-15pt}
\end{figure}

\begin{figure}[ht]
    \centering
    \includegraphics[width=\linewidth]{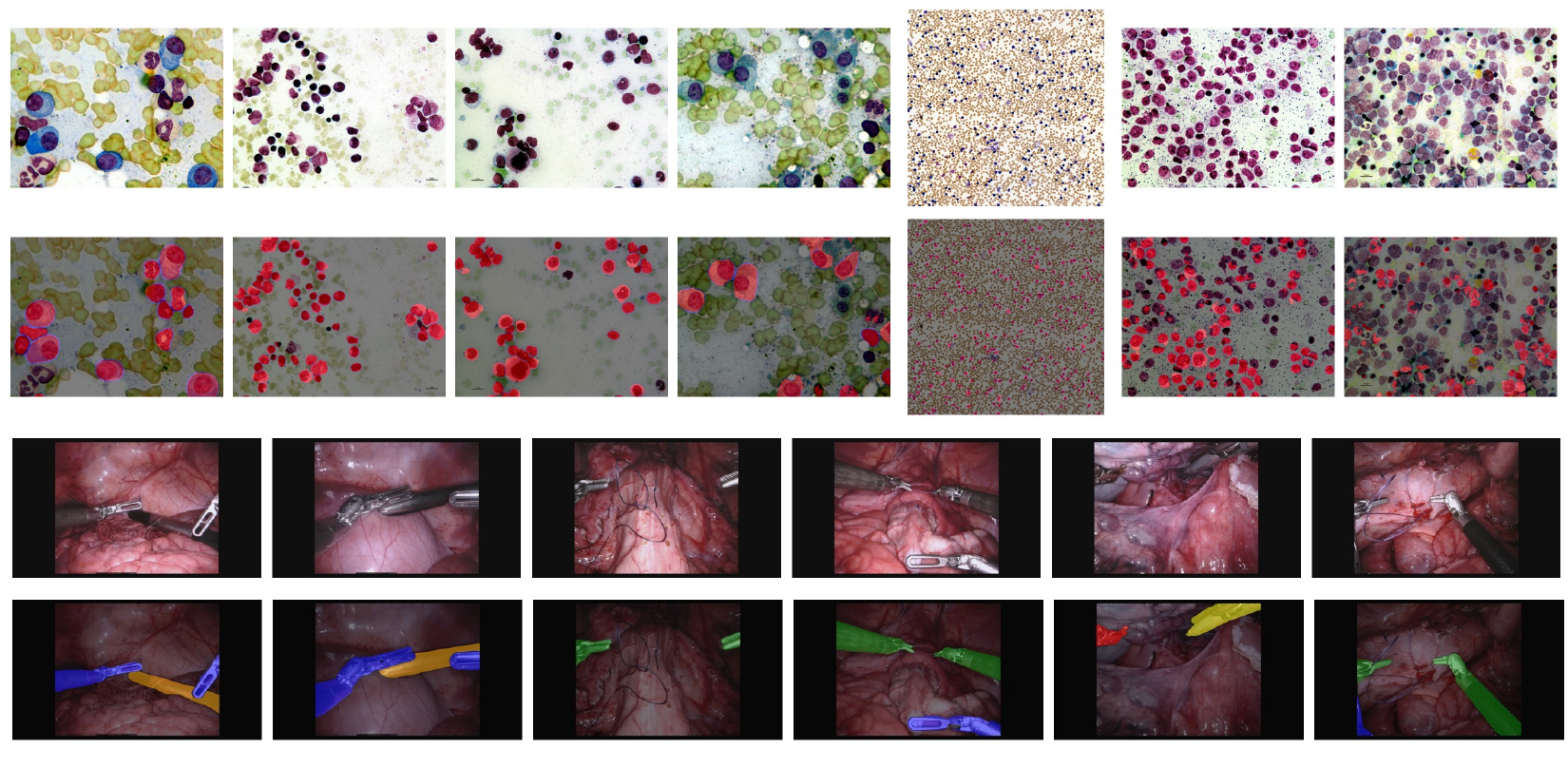}
    \vspace{-15pt}
    \caption{Visualization results of TTT-UNet on Microscopy and Endoscopy datasets. The first and second rows show the original images and TTT-UNet predictions on the Microscopy dataset, respectively. The third and fourth rows show the original images and TTT-UNet predictions on the Endoscopy dataset.}
    \label{fig: vis2d}
    \vspace{-15pt}
\end{figure}

As shown in \textbf{Figure~\ref{fig: vismir}}, the segmentation results on the Abdomen MRI dataset reveal the effectiveness of TTT-UNet in handling complex anatomical structures. The comparison between the ground truth labels and the TTT-UNet predictions indicates a strong alignment, particularly in regions with significant anatomical variability. This suggests that TTT-UNet's ability to adapt its parameters during test time enhances its precision in segmenting intricate and variable abdominal organs, making it a robust tool in scenarios where consistent accuracy is crucial. 

\textbf{Figure~\ref{fig: vis2d}} provides further insights through the segmentation results on the Endoscopy and Microscopy datasets. In the Endoscopy dataset, TTT-UNet successfully delineates the surgical instruments, which are challenging due to their small size and diverse appearances. This capability underlines the model's strength in capturing fine details and its adaptability to various shapes and textures. Similarly, in the Microscopy dataset, TTT-UNet demonstrates its robustness by accurately segmenting cell boundaries and interiors, even amidst high variability and noise levels. The model's performance in these diverse settings highlights its versatility and reliability across different medical imaging modalities.

The visual evidence presented in Figures~\ref{fig: vismir} and~\ref{fig: vis2d} aligns with the quantitative improvements reported in Table~\ref{tab:results-2d}. These results underscore TTT-UNet's consistent ability to deliver high-quality segmentations, affirming its potential as a state-of-the-art approach in medical image analysis.

\section{Discussion and conclusion}
The experimental results across multiple biomedical image segmentation tasks consistently demonstrate that TTT-UNet provides a significant improvement over state-of-the-art methods. The key factor driving this improvement is the integration of TTT layers, which allow the model to adapt to the characteristics of each test image dynamically. This capability leads to enhanced generalization, especially in tasks involving diverse and complex imaging modalities, such as 3D abdomen CT, abdomen MRI, endoscopy, and microscopy datasets.

Furthermore, TTT-UNet’s superior performance in capturing long-range dependencies and handling high anatomical variability positions it as a robust tool for clinical applications. In the case of both large-scale anatomical structures and smaller, intricate features, TTT-UNet has demonstrated the ability to adapt and deliver accurate segmentation results. This versatility is particularly crucial in clinical scenarios where precision and adaptability are essential for effective diagnosis and treatment.

One of the primary advantages of TTT-UNet lies in its capacity to dynamically adjust model parameters during the test phase, which significantly enhances segmentation accuracy. Additionally, the lower variance in performance across different datasets emphasizes the model’s robustness and consistency. However, it is important to acknowledge that the computational cost associated with test-time training could be a limitation for real-time applications. Future work should focus on optimizing the TTT layers to minimize computational overhead without compromising performance.

In conclusion, TTT-UNet represents a significant advancement in biomedical image segmentation by offering a flexible and adaptive solution. Its ability to consistently outperform other models in both 2D and 3D segmentation tasks reinforces its potential as a state-of-the-art model for medical image analysis. As the model evolves, further optimization of test-time adaptation strategies and integration with large-scale datasets will pave the way for broader clinical adoption and deployment.

\section{Acknowledgements}
This work is partially supported by the National Science Foundation grants (MRI-2215789, IIS-2319451, CRII-2246067, ATD-2427915, POSE-2346158), National Institutes of Health grants (R21EY034179), and Lehigh’s grants under CORE (001250) and FRG (S00011497).




\bibliographystyle{unsrt}
\bibliography{references}

\end{document}